\documentclass[epsfig,prd,aps,preprint,superscriptaddress]{revtex4}
\usepackage{epsfig}
\usepackage{graphics}

\begin{document}
\title{On the degrees of freedom of a black hole}
\author{Diego Pav\'on}
\email{diego.pavon@uab.es} \affiliation{Departamento de
F\'{\i}sica, Universidad Aut\'onoma de  Barcelona, Facultad de
Ci\'{e}ncias,\\ 08193 Bellaterra (Barcelona) Spain.}

\begin{abstract}
\noindent By examining whether black holes fulfill the theorem of
equipartition of energy we find that the notion of degrees of
freedom, previously introduced for cosmic horizons, is meaningful
in the case of Schwarzschild and Kerr black holes. However, for
Reissner-N\"{o}rdstrom and Kerr-Newman black holes this notion
fails.
\end{abstract}

\maketitle
\section{Introduction}
\noindent As is widely acknowledged nowadays, the complete
gravitational collapse of matter typically produces a black hole
\cite{oppenheimer, East}. Very generally, a fraction of the energy
of the pre-existing body is radiated away  during the collapse
\cite{paul} whereby the final entropy (the entropy carried in the
outflow of energy plus the entropy of the resulting black hole) is
necessarily larger than the entropy of the said body. Roughly
speaking, the latter is expected to vary with the number of
particles composing it and therefore with the number of unfrozen
degrees of freedom.
\\  \

\noindent In the case of a black hole the concept of ``number of
particles" is meaningless but the notion of ``number of unfrozen
degrees of freedom" may not. According to the equipartition
theorem each unfrozen degree of freedom of a system at equilibrium
at temperature $\, T \, $ contributes a fixed quantity, say $\xi
k_{B} \, T\, $, to the energy of the system. Since black holes
possess entropy and temperature it is natural to associate a
certain number $\, {\cal N} \, $ of (unfrozen) degrees of freedom
to them. This number is usually taken as the area of the black
hole's event horizon over the Planck length to the square, ${\cal
A}/\ell^{2}_{p}\, $ \textemdash see e.g. \cite{paddy2010}.
However, presently, this remains an unsubstantiated conjecture. To
the best of our knowledge, a proof from first principles based on
quantum gravity is still lacking.
\\  \

\noindent It is noteworthy  that while  the entropy of a gas is
approximately proportional to the number of particles composing it
\cite{Frautschi} \textemdash therefore to the mass of the
gas\textemdash, the entropy of the  black hole resulting from the
full gravitational collapse of the gas varies as the square of its
mass. Loosely speaking, this sudden and huge increase may be
regarded as a phase transition. This sharply contrasts with phase
transitions in non-gravitational physics since in the latter, this
entropy increase solely occurs when the system transits from a
condensed state to a non-condensed one. It illustrates of how
deeply affected  the thermodynamic behavior of a system is when
gravity dominates its evolution.
\\  \

\noindent In the absence of a quantum gravity argument in favor of
the said conjecture it seems worthwhile to study whether the area
(in Planck's units) of a classical black hole satisfies the
equipartition of energy. If it does,  then the notion of ``degrees
of freedom" of a black hole should receive a strong support. In
this paper we take the aforesaid conjecture (i.e., that the number
of degrees of freedom of a black hole is given by $\, {\cal N} =
{\cal A}/\ell^{2}_{p}$) as a working hypothesis and explore
whether the equipartition theorem, $M = \xi {\cal N} k_{B} T$, is
satisfied. As it turns out, Schwarzschild black holes obey it with
$\xi = 1/2$. Strictly speaking rotating black holes do not fulfil
it though they satisfy a generalized version of the theorem.
Charged black holes fail to comply with it.  We use units such
that $\, c = G = k_{B} = \hbar = 1$.

\section{Kerr and Schwarzschild black holes}
\noindent We begin by considering whether rotating, uncharged
black holes fulfill the equipartition of energy.
\\  \

\noindent The Smarr's formula \cite{smarr} in this case reads
\begin{equation}
M = \frac{\kappa}{4 \pi}\, {\cal A} \, + \, 2 \, \Omega \, J\, ,
\label{smarr2}
\end{equation}
where $\, J$ and $\, \Omega$ are the angular momentum of the black hole
and the angular velocity of the event horizon, respectively, and
\begin{equation}
\kappa = \frac{1}{2M} \, \frac{\sqrt{1 \, - \, J^{2}/M^{4}}}{1 \,
+ \, \sqrt{1 \, - \, J^{2}/M^{4}}} \label{tkbh}
\end{equation}
denotes the acceleration felt  by a test particle on the event
horizon. Related to it is the black hole temperature defined by $T =
\kappa/2\pi$.
\\   \

\noindent Using $\, J = M \Omega \, {\cal A}/4 \pi$ and
identifying the number of unfrozen degrees of freedom with the
horizon area leads to
\begin{equation}
M = \frac{\pi}{2 \pi \, - \, \Omega^{2} \, {\cal N}}\, {\cal N} T.
 \label{kerrdf}
\end{equation}
Notice that the condition $\, J^{2} < M^{4}\, $ ensures that for
regular black holes  ${\cal N} < 2 \pi/\Omega^{2}$, hence the
right hand side of last equation never becomes negative. For
extreme Kerr black holes ($J^{2} = M^{4}$) the first term on the
right diverges and $\, {\cal N} T$ vanishes. Also, the aforesaid
side remains finite and equal to $\, M$, as it should.
\\  \

\noindent Equation (\ref{kerrdf}) suggests that every degree of
freedom contributes $\, \xi T$, where
\begin{equation}
\xi = \frac{\pi}{2 \pi \, - \, \Omega^{2} {\cal N}} \, ,
\label{xi-k}
\end{equation}
to the black hole mass. Thus, strictly speaking, the equipartition
theorem does not hold for Kerr black holes, since $\xi$ depends on
${\cal N}$. Nevertheless, they satisfy a direct generalization of
the theorem since, given ${\cal N}$, each degree of freedom
contributes to $\, M$ by the same amount.
\\   \

\noindent Consider two black holes of the same area but with different
angular velocities  (and therefore, different masses and
temperatures). The black hole with the larger $\, \Omega \,$ will have the
larger $\, \xi$ and, because of (\ref{tkbh}) and the relationship
$\, J = M \Omega \, {\cal A}/4 \pi$, the lower temperature. This
is most reasonable from the point of view of the equipartition
theorem.
\\   \

\noindent On the other hand, the fact that $\, \xi$ remains finite
for non-extreme Kerr black holes is fully consistent with the
third law of black hole thermodynamics (i.e., that $\, \kappa$
cannot be made vanish by a continuous process of absorption of
matter that satisfies the weak energy condition  \cite{werner}).
We may conclude that the third law and the non-divergence of $\,
\xi$  mutually imply  each other.
\\   \

\noindent Further, this is in keeping: (i) with the well known
fact that in the absorption of a particle the variation of the black hole
parameters is constrained by the relationship $\, \delta J <
\delta M/\Omega$, where $ \, \delta J$ and $\delta M$ coincide
with the angular momentum and energy, respectively, of the
particle measured by  an observer at infinity -see, e.g.
\cite{wald1984}. And (ii), with Page result that in Hawking
radiance the angular momentum of the black hole is emitted faster than its
energy \cite{page1976}.
\\  \

\noindent Clearly, one can write the quantity $\, \xi$ in terms of
$\, J$ and $\, M$ but the simple formula (\ref{kerrdf}) for the
mass gets lost. Instead, one obtains a cubic equation for $M$, not
an expression of the equipartition theorem since the mass does not
longer appear proportional to the temperature when $\Omega$ is
replaced by $J$. To see this from a different angle, let us assume
that (thanks to astrophysical measurements) the area, angular
velocity and temperature of a Kerr black hole are experimentally
known, but neither its angular momentum nor its mass (though they
can be derived). Then, while $M$ can be obtained directly by
(\ref{kerrdf}) it cannot by the Smarr's formula, Eq.
(\ref{smarr2}).
\\   \

\noindent From (\ref{xi-k}) it is immediately seen that for
Schwarzschild black holes ($J = 0$) the dimensionless quantity $\,
\xi$ reduces to a constant ($1/2$ in this case). Therefore, we can
say that non-rotating, uncharged black holes satisfy the
equipartition theorem. Intriguing enough, this $\xi$ value
coincides with the corresponding one to  systems whose Hamiltonian
is a quadratic function of the linear momentum of its particles
\textemdash see e.g. \cite{huang}; something far removed from
fully gravitationally collapsed objects.
\section{Kerr-Newman black holes}
\noindent In the case of rotating  charged black holes  Smarr's
formula  generalizes to
\begin{equation}
M = \frac{\kappa}{4 \pi}\, {\cal A} \, + \, 2 \Omega \, J \, + \,
\Phi \, Q \, , \label{smarr5}
\end{equation}
where
\begin{equation}
\Phi = \frac{1}{M}\, \left[\frac{Q}{2}\, + \, \frac{2 \pi
Q^{3}}{{\cal A}}\right]
 \label{potential}
\end{equation}
stands for the electrostatic potential on the black hole event horizon generated by the charge $\, Q$.
\\  \

\noindent In view of this it is  not possible to express the black
hole mass as $\, M = \xi {\cal N} T$ being $\, \xi$  a function of
$\, J$ and $\, Q$ but not of $\, M$. This implies that the
equipartition theorem does not hold for charged black holes and,
therefore, they must possess additional degrees of freedom whose
contribution to the black hole mass do not obey the simple $\, \xi
T$ rule. Hence a baffling situation arises. Considers a Kerr black
hole. There, as we have seen, a generalized version of the
equipartition theorem is satisfied. However, it suffices the fall
of a single electron on the black hole for the said theorem to
break down right away.
\\   \

\noindent One may try to solve this puzzle as follows. When the
charge is small ($ Q^{2} \ll M^{2}$), the Smarr's formula can be
approximated by
\begin{equation}
M \simeq \frac{\kappa}{4 \pi} \, {\cal A} \, + \, \frac{\Omega^{2} \, {\cal A}}{2 \pi}  M \, + \, \frac{Q^{2}}{2M}\, ,
\label{approximated3}
\end{equation}
where we have used the relationship $\, J = M \Omega  {\cal A}/4 \pi$. Solving for $\, M$, discarding the minus sign before the
square root and expanding the latter in terms of $\, Q^{2}$, we arrive to $M \simeq \xi {\cal N} T$ with
\begin{equation}
\xi = \frac{\pi}{2 \pi \, - \, \Omega^{2} {\cal N}} \, \left[1 \, +  \,
\frac{2 \pi - \Omega^{2} {\cal N}}{4 \, \pi \left( \frac{\kappa {\cal N}}{4 \pi}\right)^{2}} \, Q^{2} \right] \, + \,
{\cal O}(Q^{4}).
\label{xikQ}
\end{equation}
Nevertheless, this does not solve the problem at all because  the
black hole temperature enters (via $\kappa$) the expression for
$\xi$. Therefore, the generalized equipartition theorem fails also
when the electric charge is small; i.e., it does not cease to hold
smoothly but abruptly and the puzzle remains.
\\  \

\noindent Altogether, the fulfillment of the equipartition theorem
by Schwarzschild and a generalized version of it for Kerr black
holes strongly suggests that the notion of degrees of freedom
makes sense for these black holes and that the area of their
horizon gives (in Planck units) the number of their unfrozen
degrees freedom. For charged black holes, however,  the theorem
breaks down and the black hole area does no longer counts the
aforesaid number.
\section{Discussion}
\noindent In the absence of a proof from first principles of the
conjecture that the number of degrees of freedom of a black hole
is given by the area (in Planck units)  of its event horizon, it
seems reasonable to explore  whether the said number is consistent
with the equipartition theorem of statistical physics. A positive
answer would lend support to that conjecture.  However, the
overall result is inconclusive. While Schwarzschild black holes
fulfill the theorem and Kerr black holes satisfy a generalized
version of it, charged black holes do not.
\\  \

\noindent As we have seen in the second section, there is a strong
connection between $\,\xi$ and $\, \kappa$. The natural
requirement that the former should remain finite implies that the
latter cannot vanish (and viceversa). So, in a way, the third law
of black hole thermodynamics \cite{werner} sets an upper limit,
for a given angular momentum, $\, J$, on $\, \xi$. To the best of
our knowledge, this interesting feature was never noticed.
\\   \

\noindent One cannot avoid  wonder why neither
Reissner-Nordstr\"{o}m nor Kerr-Newman black holes comply with the
equipartition theorem.  Why the electric charge behaves so
dissimilarly to the angular momentum in this respect? It may be
related to the fact that in Schwarzschild spacetimes there is one
killing vector, $t^{a} =
\partial x^{a}/\partial t$, directly connected to existence of the
black hole mass. In the case of Kerr spacetimes there is one
further Killing vector entirely related to the rotation of the
event horizon, i.e., $\phi^{a} = \partial x^{a}/\partial \phi$. By
contrast, no Killing vector is related to the presence of the
electric charge, neither in Reissner-Nordstr\"{o}m nor Kerr-Newman
spacetimes. Nevertheless, for the time being the relationship, if
it exists, between the mentioned Killing fields and the
equipartition theorem remains a mystery. Hopefully, quantum
gravity will provide someday a convincing answer to it.
\\   \

\noindent One may argue that, actually, the equipartition theorem
is confined to equilibrium systems and, strictly speaking, this is
not the case of isolated black holes  since their mass steadily
diminish via Hawking emission. We do not think this  is a serious
hurdle. The mass of a classical black hole is much larger than the
Planck mass. Hence the rate of mass loss is negligible even when
compared to the Hubble constant (bear in mind that $\, - \dot{M}
\sim M^{-2}$). On the other hand, a Kerr black hole can be brought
to a stable thermodynamic equilibrium by enclosing it in a box
filled with radiation at the temperature of the black hole and
rotating at the  angular speed of the latter, provided that the
radiation energy in the box does not exceed one fourth of the
black hole mass \cite{hawking1976}.
\\  \

\noindent Thus far our interest was restricted to classical black
holes. When quantization is incorporated the equipartition theorem
still holds, provided that $M \gg 1$, but the number of degrees of
freedom differs though slightly. Recalling that the area spectrum
of quantum Schwarzschild black holes is \cite{prl-hod}
\begin{equation}
{\cal A}_{n} = 4 (\ln 3) \, n  \qquad (n = 1, 2, ...),
\label{asqs}
\end{equation}
and that $ \, {\cal N}$ has to be an integer, it follows $\,   {\cal N} =
(\ln 3)\, {\cal A}_{n}$. Thereby,
\begin{equation}
M =  \frac{\ln 3}{2}\, {\cal N} \, T \label{mqsbh}
\end{equation}
valid for $n \gg 1$. Thus,  each degree of freedom contributes $\,
(\ln 3) T/2$   to the mass of a large Schwarzschild quantum black
hole.

\section*{Acknowledgments}
\noindent It is a pleasure to thank Juan Camacho, Jos\'{e} Pedro
Mimoso, Diego Rubiera-Garc\'{i}a and Bin Wang for useful
conversations on the subject of this work.

\end{document}